\documentclass[final,3p,times]{elsarticle}

\biboptions{sort&compress}

\usepackage{euscript}
\usepackage{graphicx}
\usepackage{amssymb}
\usepackage{amsmath}
\usepackage{amsfonts}

\newcommand{\Rea}{\mathop{\mathrm{Re}}\nolimits}
\newcommand{\Ima}{\mathop{\mathrm{Im}}\nolimits}
\newcommand{\ci}{\mathop{\mathrm{ci}}\nolimits}
\newcommand{\si}{\mathop{\mathrm{si}}\nolimits}

\journal{Nuclear Instruments and Methods in Phys. Res. B}

\begin{document}

\begin{frontmatter}

\title{Second-order Born approximation for the scattering phase shifts:
Application to the Friedel sum rule}

\author[label1,label2]{Hrachya~B.~Nersisyan\corref{cor1}}
\ead{hrachya@irphe.am}

\author[label3]{Jos\'{e} M. Fern\'{a}ndez-Varea}
\ead{jose@ecm.ub.edu}

\address[label1]{Plasma Theory Group, Institute of Radiophysics and Electronics,
0203 Ashtarak, Armenia}

\address[label2]{Centre of Strong Fields Physics, Yerevan State University,
Alex Manoogian str.\ 1, 0025 Yerevan, Armenia}

\address[label3]{Facultat de F\'{\i}sica (ECM and ICC), Universitat de Barcelona,
Diagonal 645, E-08028 Barcelona, Spain}

\cortext[cor1]{Corresponding author}

\begin{abstract}

Screening effects are important to understand various aspects of
ion-solid interactions and, in particular, play a crucial role in the
stopping of ions in solids. In this paper the phase shifts and
scattering amplitudes for the quantum-mechanical elastic scattering
within up to the second-order Born (B2) approximation are revisited for
an arbitrary spherically-symmetric electron-ion interaction potential.
The B2 phase shifts and scattering amplitudes are then used to derive
the Friedel sum rule (FSR) involving the second-order Born corrections.
This results in a simple equation for the B2 perturbative screening
parameter of an impurity ion immersed in a fully degenerate electron gas
which, as expected, turns out to depend on the ion atomic number $Z_{1}$
unlike the first-order Born (B1) screening parameter reported earlier by
some authors. Furthermore, our analytical results for the Yukawa,
hydrogenic, Hulth\'{e}n, and Mensing potentials are compared, for both
positive and negative ions and a wide range of one-electron radii, to
the exact screening parameters calculated self-consistently by imposing
the FSR requirement. It is shown that the B2 screening parameters agree
excellently with the exact values at large and moderate densities of the
degenerate electron gas, while at lower densities they progressively
deviate from the exact numerical solutions but are nevertheless more
accurate than the prediction of the B1 approximation. In addition, a
simple Pad\'{e} approximant to the Born series has been developed that
improves the performance of the perturbative FSR for any negative ion as
well as for $Z_{1}=+1$.

\end{abstract}


\end{frontmatter}


\section{Introduction}
\label{sec:int}

The problem of ion interactions in condensed matter continues to be the
subject of intense experimental and theoretical research. These
interactions are relevant to understand, among others, the behavior of
static impurities in metals, such as the resistivity of impurities and
metallic solutions \cite{mot58}, or the stopping of ions in solids
\cite{kum81}. The screening of the intruder ion in the host medium plays
a key role in these phenomena.

A number of approaches, both perturbative and non-perturbative, have
been devised over the years to describe the basic processes of ion-solid
interactions. In particular, following the pioneering works of Lindhard
\cite{lin54}, and Lindhard and Winther \cite{lin64}, many calculations
have been done within the framework of linear-response theory (see,
e.g., \cite{abr98,zwi99,ner00,ner02,ner03,ner04,her05,ner08,ner09}),
which enables a unified description of dynamical screening, plasmon
excitation, and creation of electron-hole pairs. Most of these
calculations are based on the dielectric function in the random-phase
approximation (RPA) which is valid in the weak-coupling (i.e.,
high-density) limit of a degenerate electron gas (DEG). The main
shortcoming of linear-response theory is observed in the low-velocity
limit since the interaction effects become too strong to be properly
accounted for by perturbative approximations \cite{ech91}.

Non-perturbative (i.e., non-linear) methods also provide a reasonable
description of ion-solid interactions. For instance, the kinetic theory
\cite{sig82,fer84} involves the transport cross section for
dynamically-screened interactions including quantum effects in the whole
velocity range. However, though non-perturbative in the ion-solid
coupling, the formalism does not include contributions from collective
(plasmon) excitations to the stopping power, which are important in the
high-velocity regime when the interaction potential becomes highly
anisotropic. These quantum-mechanical models were initially proposed for
the case of slow ions. In~\cite{bri74} a transport cross section
approach based on the partial-wave expansion was introduced to calculate
the stopping power of positive ions in channeling conditions, explaining
qualitatively the observed oscillatory behavior of this quantity with
the ion atomic number $Z_{1}$ (``$Z_{1}$ oscillations''). A more
rigorous many-body representation of the non-linear screening and
stopping processes in a homogeneous DEG was subsequently given in
\cite{ech81,ech86,ech89,gra91} working within density-functional theory.
A computationally convenient simplification is achieved if the numerical
density-functional-theory potential is substituted by an analytical
electron-ion interaction potential with a free parameter that is
adjusted self-consistently requiring that the scattering phase shifts
satisfy the Friedel sum rule \cite{mei75,fer77,apa87,ven88,sor90,nag94}.
It is possible to introduce in the analytical potential more parameters,
which are adjusted demanding the fulfillment of additional constraints
like Kato's cusp condition in the self-consistent procedure
\cite{apa88,fig10}. The aforementioned non-linear approaches can be
adapted immediately to deal with inhomogeneous DEGs having recourse to
the local-plasma approximation; this allows a realistic description of
the screening and energy loss of low-energy ions in the
spatially-varying electron densities encountered in solids
\cite{cal93,cal94,wan97,fig10}.

The problem of extending the quantum treatments to finite
velocities has been addressed more recently in the context of
density-functional theory \cite{zar95} and by means of model potentials
with and without the Born approximation
\cite{nag96,lif98,nag98,ari02,ner05}. For instance, extensions of the
Friedel sum rule to finite velocities have been formulated either making
use of the first-order Born (B1) approximation~\cite{nag96} or including
all orders in the interaction strength~\cite{lif98}. In these
formalisms the dynamical potential is replaced by a spherically
symmetric one which facilitates the application of the conventional
partial-wave analysis of one-electron scattering phase shifts. In the
present article we too rely on this assumption.

This work concerns itself with the perturbative treatment of screening
effects in the case of static or slow ions in a DEG. The
B1 approximation yields a screening parameter that is independent
of the ion atomic number $Z_{1}$ and hence is the same for slow
particles and antiparticles \cite{ech89}. This situation is somewhat
unsatisfactory in analyzing the available experimental data on proton
and antiproton energy losses in various solids \cite{mol02,mol04,mol08}.
Recently, using the Friedel sum rule the screening lengths within the
second-order Born (B2) approximation have been deduced in~\cite{ner05}
for the Yukawa and Mensing interaction potentials. These B2 screening
lengths pertaining to protons and antiprotons agree satisfactorily with
the exact numerical solutions at electron densities typical of metals.
However, in~\cite{ner05} only some simplified expressions of the
screening lengths were studied and further investigation on this topic
seemed desirable. To carry out this idea we evaluate the scattering
amplitude and phase shifts within the B2 approximation and for an
arbitrary interaction potential, which allows us to formulate explicitly
the Friedel sum rule at the same level of the B2 approximation.

The scattering phase shifts are derived within the B2 approximation in
Section~\ref{sec:2} for an arbitrary spherically-symmetric interaction
potential. In Section~\ref{sec:3} we employ these results to derive the
second-order Friedel sum rule. Based on this sum rule we have developed
a simple but general equation that determines the screening length,
within the B2 approximation, for arbitrary screened potentials.
Moreover, this equation has been specified for the Yukawa, hydrogenic,
Hulth\'{e}n, and Mensing potentials. The perturbative results for these
potentials are compared, in Section~\ref{sec:4}, with the corresponding
exact solutions calculated from the Friedel sum rule in a wide range of
DEG densities and for several charges of the impurity ion. Finally, the
Pad\'{e} approximant for the obtained Born series in the Friedel sum
rule has also been examined. Some details of the analytical calculations
are included in Appendices~A, B, and C.


\section{First- and second-order phase shifts and scattering amplitudes}
\label{sec:2}

In this section we deduce, within the Born approximation, the B1 and B2
scattering phase shifts using the systematic perturbative expansion of
the exact scattering amplitude and the exact electronic wave function at
the final state (after scattering). Although the B1 and B2 phase shifts
are well known (see, e.g., \cite{joa75}), we suggest an alternative
derivation of these quantities which is more suitable for the evaluation
of the perturbative screening parameters in Section~\ref{sec:3}. The
starting point is the exact relation between the scattering amplitude
$f(k,\theta)$ for the elastic scattering and the phase shifts
$\delta_{\ell}(k)$ which is given by \cite{lan81}
\begin{equation}
f(k,\theta) =
\frac{1}{2\mathrm{i}k} \, \sum_{\ell=0}^{\infty} (2\ell+1)
\left( \mathrm{e}^{2\mathrm{i}\delta_{\ell}} - 1 \right)
P_{\ell}(\cos\theta).
\label{eq:f_exact}
\end{equation}
Here $k$ is the electron wave number, $\theta$ is the scattering angle,
and $P_{\ell}$ are the Legendre polynomials. Within the Born
approximation we assume that the $n$-th order phase shifts are determined
by $\delta_{\ell,\mathrm{B}n}\sim B^{n}$ with $n=1,2,\ldots$, where $B$
is the Born ``smallness parameter'' which should be precisely determined
for each interaction potential. It is clear that $B\sim V(r)$, where
$V(r)$ is the isotropic (i.e., spherically symmetric) interaction
potential of the colliding particles. In this paper we restrict
ourselves to the B2 approximation and look for the phase shifts in a
perturbative manner
$\delta_{\ell}=\delta_{\ell,\mathrm{B}1}+\delta_{\ell,\mathrm{B}2}$,
where $\delta_{\ell,\mathrm{B}1}$ and $\delta_{\ell,\mathrm{B}2}$ are
the first- and second-order phase shifts, respectively. Analogously, the
scattering amplitude in Eq.~(\ref{eq:f_exact}) can be written as
$f=f_{\mathrm{B}1}+f_{\mathrm{B}2}$, where
$f_{\mathrm{B}1}$ and $f_{\mathrm{B}2}$ are the first-
and second-order amplitudes, respectively. Next, using the perturbative
expansion of the phase shift we rewrite the exponential factor in
Eq.~(\ref{eq:f_exact}) in the form
$(\mathrm{e}^{2\mathrm{i}\delta_{\ell}}-1)/2\mathrm{i}\simeq
\delta_{\ell}+\mathrm{i}\delta_{\ell}^{2}\simeq
\delta_{\ell,\mathrm{B}1}+\delta_{\ell,\mathrm{B}2}
+\mathrm{i}\delta_{\ell,\mathrm{B}1}^{2}$. Then, keeping only terms up
to the second order we have
\begin{eqnarray}
f_{\mathrm{B}1}(k,\theta)
& = &
\frac{1}{k} \,
\sum_{\ell=0}^{\infty} (2\ell+1) \,
\delta_{\ell,\mathrm{B}1} \, P_{\ell}(\cos\theta),
\label{eq:f_B1_a}
\\
f_{\mathrm{B}2}(k,\theta)
& = &
\frac{1}{k} \,
\sum_{\ell=0}^{\infty} (2\ell+1)
\left( \delta_{\ell,\mathrm{B}2} + \mathrm{i}\delta_{\ell,\mathrm{B}1}^{2} \right)
P_{\ell}(\cos\theta).
\label{eq:f_B2_a}
\end{eqnarray}
Note that the imaginary part of $f_{\mathrm{B}1}(k,\theta)$ vanishes
while the imaginary part of the B2 scattering amplitude at $\theta=0$
(forward scattering) satisfies the optical theorem
$\Ima\big[f_{\mathrm{B}2}(k,0)\big]=(k/4\pi)\,\sigma_{\mathrm{tot,B}1}(k)$
\cite{joa75} within the B2 approximation, where
$\sigma_{\mathrm{tot,B}1}(k)$ is the B1 total elastic cross section.

To determine the first- and second-order phase shifts,
Eqs.~(\ref{eq:f_B1_a}) and (\ref{eq:f_B2_a}) should be compared with the
corresponding scattering amplitudes extracted from the systematic
perturbative expansion of the exact electronic wave function at the
final state (see, e.g., \cite{lan81}). This procedure is straightforward
and leads to
\begin{eqnarray}
f_{\mathrm{B}1}(k,\theta)
& = &
-\frac{m_{\mathrm{e}}}{2\pi\hbar^{2}}
\int V(r) \,
\varphi_{\mathbf{k}_{f}}^{\ast}(\mathbf{r}) \,
\varphi_{\mathbf{k}_{i}}(\mathbf{r}) \,
\mathrm{d}\mathbf{r},
\label{eq:f_B1_b}
\\
f_{\mathrm{B}2}(k,\theta)
& = &
\left( \frac{m_{\mathrm{e}}}{2\pi\hbar^{2}} \right)^{2}
\int \! \int
\frac{\mathrm{e}^{\mathrm{i}k\left\vert\mathbf{r}-\mathbf{r}'\right\vert}}
     {\left\vert \mathbf{r}-\mathbf{r}'\right\vert} \,
V(r) \, V(r') \,
\varphi_{\mathbf{k}_{f}}^{\ast}(\mathbf{r}) \,
\varphi_{\mathbf{k}_{i}}(\mathbf{r}') \,
\mathrm{d}\mathbf{r} \, \mathrm{d}\mathbf{r}',
\label{eq:f_B2_b}
\end{eqnarray}
where $\mathbf{k}_{i}$ and $\mathbf{k}_{f}$ are the initial and final
wave vectors of the electron, respectively. Let us recall that we assume
here an elastic scattering process with momentum conservation, i.e.,
$k_{f}=k_{i}=k$. Also
$\varphi_{\mathbf{k}}(\mathbf{r})=\mathrm{e}^{\mathrm{i}\mathbf{k}\cdot\mathbf{r}}$
is the unperturbed electronic wave function corresponding to the wave
vector $\mathbf{k}$. Thus, from Eqs.~(\ref{eq:f_B1_b}) and
(\ref{eq:f_B2_b}) one has
\begin{eqnarray}
f_{\mathrm{B}1}(k,\theta)
& = &
-\frac{m_{\mathrm{e}}}{2\pi\hbar^{2}} \, \widetilde{V}(q),
\label{eq:f_B1_c}
\\
f_{\mathrm{B}2}(k,\theta)
& = &
\left( \frac{m_{\mathrm{e}}}{2\pi\hbar^{2}} \right)^{2}
\int \! \int
\frac{\mathrm{e}^{\mathrm{i}k\left\vert\mathbf{r}-\mathbf{r}'\right\vert}}
     {\left\vert\mathbf{r}-\mathbf{r}'\right\vert} \,
V(r) \, V(r') \,
\mathrm{e}^{-\mathrm{i}\mathbf{k}_{f}\cdot\mathbf{r}} \,
\mathrm{e}^{\mathrm{i}\mathbf{k}_{i}\cdot\mathbf{r}'} \,
\mathrm{d}\mathbf{r} \, \mathrm{d}\mathbf{r}'.
\label{eq:f_B2_c}
\end{eqnarray}
Here $\mathbf{q}=\mathbf{k}_{f}-\mathbf{k}_{i}$ is the momentum transfer
in the elastic collision with $q=2k\sin(\theta/2)$ and
$\widetilde{V}(q)$ is the Fourier transform of the interaction
potential. Note that $\widetilde{V}(q)$ is spherically symmetric in
momentum space and is given by
\begin{equation}
\widetilde{V}(q) =
\int_{0}^{\infty} V(r) \, j_{0}(qr) \, 4\pi r^{2} \, \mathrm{d}r,
\label{eq:V_Fourier}
\end{equation}
where $j_{\ell}(z)$ are the spherical Bessel functions of the first kind
and order $\ell$ with $j_{0}(z)=\sin z/z$ \cite{abr72,gra80}.

Equation~(\ref{eq:f_B1_c}) can be developed further if in
Eq.~(\ref{eq:V_Fourier}) we replace $j_{0}(qr)$ with (see Eq.~(10.1.45)
in~\cite{abr72})
\begin{equation}
j_{0}\big(2kr\sin(\theta/2)\big) =
\sum_{\ell=0}^{\infty} (2\ell+1) \, j_{\ell}^{2}(kr) \, P_{\ell}(\cos\theta).
\label{eq:9}
\end{equation}
Comparing the resulting expression for $f_{\mathrm{B}1}(k,\theta)$ with
Eq.~(\ref{eq:f_B1_a}) we conclude that (see, e.g., \cite{joa75})
\begin{equation}
\delta_{\ell,\mathrm{B}1}(k) =
-\frac{2m_{\mathrm{e}}k}{\hbar^{2}} \,
\int_{0}^{\infty} V(r) \, j_{\ell}^{2}(kr) \, r^{2} \, \mathrm{d}r.
\label{eq:10}
\end{equation}
On the other hand, in Eq.~(\ref{eq:f_B2_c}) we may expand
$\mathrm{e}^{\mathrm{i}k|\mathbf{r}-\mathbf{r}'|}/|\mathbf{r}-\mathbf{r}'|$
with the help of Eq.~(B.43) in~\cite{joa75} and use the Rayleigh
expansion of the plane wave over spherical harmonics to deal with
$\mathrm{e}^{-\mathrm{i}\mathbf{k}_{f}\cdot\mathbf{r}}$ and
$\mathrm{e}^{\mathrm{i}\mathbf{k}_{i}\cdot\mathbf{r}'}$. After some
algebraic manipulations that involve the orthonormality relation and the
addition theorem of spherical harmonics \cite{gra80,abr72} one obtains
\begin{eqnarray}
f_{\mathrm{B}2}(k,\theta)
& = &
\mathrm{i}k \left( \frac{2m_{\mathrm{e}}}{\hbar^{2}} \right)^{2}
\sum_{\ell=0}^{\infty} (2\ell+1) \, P_{\ell}(\cos\theta)
\nonumber
\\
& &
\mbox{}
\times \int_{0}^{\infty} V(r) \, j_{\ell}(kr) \, r^{2} \, \mathrm{d}r
\nonumber
\\
& &
\mbox{}
\times \int_{0}^{\infty} V(r') \,
j_{\ell}(kr_{<}) \, h_{\ell}^{(1)}(kr_{>}) \, j_{\ell}(kr') \, r'^{2} \,
\mathrm{d}r'.
\label{eq:11}
\end{eqnarray}
Here $h_{\ell}^{(1)}(z)=j_{\ell}(z)+\mathrm{i}n_{\ell}(z)$ are the
spherical Hankel functions of the first kind, $n_{\ell}(z)$ are the
spherical Bessel functions of the second kind \cite{abr72},
$r_{<}=\min\{r,r'\}$, and $r_{>}=\max\{r,r'\}$. Comparing now the real
parts of Eqs.~(\ref{eq:11}) and (\ref{eq:f_B2_a}) we finally get (see,
e.g., \cite{joa75})
\begin{eqnarray}
\delta_{\ell,\mathrm{B}2}(k)
& = &
-\left( \frac{2m_{\mathrm{e}}k}{\hbar^{2}} \right)^{2}
\int_{0}^{\infty} V(r) \, j_{\ell}(kr) \, r^{2} \, \mathrm{d}r
\nonumber
\\
& &
\mbox{}
\times \int_{0}^{\infty} V(r') \,
j_{\ell}(kr_{<}) \, n_{\ell}(kr_{>}) \, j_{\ell}(kr') \, r'^{2} \, \mathrm{d}r'.
\label{eq:12}
\end{eqnarray}

It is easy to verify that the imaginary parts of Eqs.~(\ref{eq:11}) and
(\ref{eq:f_B2_a}) are identical because of the relation (\ref{eq:10})
for the B1 phase shifts. Equations (\ref{eq:10}) and (\ref{eq:12}) thus
represent the first- and second-order phase shifts, respectively. As
expected, they are proportional to $\sim V(r)$ and $\sim V^{2}(r)$,
respectively. It should be emphasized that Eqs.~(\ref{eq:10}) and
(\ref{eq:12}) are valid when $B<1$, which in particular is satisfied at
high velocities (large $k$) or in the case of a weak interaction
potential. In addition, the validity of the Born approximation requires
that $\delta_{\ell,\mathrm{B}2}<\delta_{\ell,\mathrm{B}1}$, which in
general is fulfilled at high velocities. At small velocities (small $k$)
using the asymptotic behavior of the spherical Bessel functions
\cite{abr72} it is not difficult to show that the B1 and B2 phase shifts
of Eqs.~(\ref{eq:10}) and (\ref{eq:12}) behave as $\delta_{\ell}\sim
(k\lambda)^{2\ell+1}$, where $\lambda$ is the characteristic range of
the interaction potential (see, e.g., \cite{joa75,lan81}). In this case
$\delta_{\ell,\mathrm{B}2}/\delta_{\ell,\mathrm{B}1}\simeq
\overline{V}(m_{\mathrm{e}}\lambda^{2}/\hbar^{2})[2^{2\ell}(2\ell+1)]^{-1}$
is independent of $k$, but depends on the average interaction potential
$\overline{V}$ (potential at the distance $\lambda$). Therefore, at
$\overline{V}<\hbar^{2}/m_{\mathrm{e}}\lambda^{2}$ systematically
$\delta_{\ell,\mathrm{B}2}<\delta_{\ell,\mathrm{B}1}$ for all $\ell$,
while at $\overline{V}>\hbar^{2}/m_{\mathrm{e}}\lambda^{2}$ the
requirement of the Born approximation is satisfied at large angular
momentum $\ell$.

We would like to close this section with the following remark. An
alternative, shorter way to derive the first- and second-order phase
shifts within the Born approximation starts from the Calogero equation
\cite{cal67}
\begin{eqnarray}
\frac{\mathrm{d}\Delta_{\ell}(r)}{\mathrm{d}r}
& = &
-\frac{2m_{\mathrm{e}}k}{\hbar^{2}} \, r^{2} \, V(r)
\nonumber
\\
& &
\mbox{}
\times \Big[
j_{\ell}(kr)\,\cos\Delta_{\ell}(r) - n_{\ell}(kr)\,\sin\Delta_{\ell}(r)
\Big]^{2}
\label{eq:Calogero}
\end{eqnarray}
for the generalized (coordinate dependent) phase shifts
$\Delta_{\ell}(r)$. The scattering phase shifts $\delta_{\ell}$ are then
deduced according to the limit
$\delta_{\ell}=\Delta_{\ell}(r)\vert_{r\to\infty}$.
Let us note that this relation along with Eq.~(\ref{eq:Calogero})
determines the absolute phase shifts of the scattering problem while
most of the existing numerical methods deliver, instead, relative phase
shifts by solving the radial Schr\"{o}dinger equation. However, absolute
$\delta_{\ell}$ are needed for certain applications like, for example,
in the Friedel sum rule (see Eq.~(\ref{eq:FSR_exact}) below).

Let us look for the solution of Eq.~(\ref{eq:Calogero}) in a
perturbative manner. Within the B1 and B2 approximations the generalized
phase shifts are denoted as $\Delta_{\ell,\mathrm{B}1}(r)$ and
$\Delta_{\ell,\mathrm{B}2}(r)$, respectively, and from
Eq.~(\ref{eq:Calogero}) it is straightforward to see that
\begin{equation}
\Delta_{\ell,\mathrm{B}1}(r) =
-\frac{2m_{\mathrm{e}}k}{\hbar^{2}} \,
\int_{0}^{r} V(r') \, j_{\ell}^{2}(kr') \, r'^{2} \, \mathrm{d}r'
\label{eq:14}
\end{equation}
and
\begin{eqnarray}
\Delta_{\ell,\mathrm{B}2}(r)
& = &
\frac{4m_{\mathrm{e}}k}{\hbar^{2}}
\int_{0}^{r} V(r') \, j_{\ell}(kr') \, n_{\ell}(kr') \,
\Delta_{\ell,\mathrm{B}1}(r') r'^{2} \, \mathrm{d}r'
\nonumber
\\
& = &
-2 \left( \frac{2m_{\mathrm{e}}k}{\hbar^{2}} \right)^{2}
\int_{0}^{r} V(r') j_{\ell}(kr') \, n_{\ell}(kr') \, r'^{2} \, \mathrm{d}r'
\nonumber
\\
& &
\mbox{}
\times \int_{0}^{r'} V(r'') \, j_{\ell}^{2}(kr'') \, r''^{2} \, \mathrm{d}r''.
\label{eq:15}
\end{eqnarray}
Taking the limit $r\to\infty$ in~(\ref{eq:14}) and (\ref{eq:15}) we
recover Eq.~(\ref{eq:10}) for the first-order phase shifts
$\delta_{\ell,\mathrm{B}1}$ while Eq.~(\ref{eq:15}) yields
\begin{eqnarray}
\delta_{\ell,\mathrm{B}2}
& = &
-2 \left( \frac{2m_{\mathrm{e}}k}{\hbar^{2}} \right)^{2}
\int_{0}^{\infty} V(r) \, j_{\ell}(kr) \, n_{\ell}(kr) \, r^{2} \, \mathrm{d}r
\nonumber
\\
& &
\mbox{}
\times \int_{0}^{r} V(r') \, j_{\ell}^{2}(kr') \, r'^{2} \, \mathrm{d}r'.
\label{eq:16}
\end{eqnarray}
To prove the identity of Eqs.~(\ref{eq:12}) and (\ref{eq:16}) we write
the latter in an equivalent form by changing the orders of the
integrations. Then, taking half of the sum of the obtained expression
and Eq.~(\ref{eq:16}) we arrive at Eq.~(\ref{eq:12}) for the
second-order phase shifts $\delta_{\ell,\mathrm{B}2}$ derived directly
from the systematic perturbative expansion of the exact scattering
amplitude.


\section{Application to the Friedel sum rule}
\label{sec:3}

With the perturbative formalism presented in Section~\ref{sec:2}, we now
take up the main topic of this paper, namely to study the Friedel sum
rule (FSR) for a point-like static (or slow) ion in a DEG of density
$n_{\mathrm{e}}$ within up to the B2 approximation. The FSR is very
useful to adjust in a self-consistent way the electron-ion interaction
potential and the related screening length. The Born approximation has
been used before in conjunction with the FSR, but only within the first
order \cite{ech89,nag96,lif98}. This is somewhat unsatisfactory since
the resulting (first-order) screening length is independent of $Z_{1}$
and is therefore identical for particles and their antiparticles. Going
beyond the B1 approximation, and considering the B2 approximation,
allows more physical insight and furnishes useful numerical estimates of
the influence of the ion charge on the screening length in a DEG. The
problem is closely related to the formulation of the theory of the
higher-order stopping power involving the Barkas correction (see, e.g.,
\cite{zwi99} and references therein) which is $\sim Z_{1}^{3}$ and is
not symmetric with respect to the sign of $Z_{1}$.

Let us consider the usual treatment of the FSR for static ions
\cite{fri52,fri54}. It may be shown that each scattered electron
contributes to the accumulation of screening charge by an amount that,
in a partial-wave expansion, is given by the derivative of the phase
shift in the form $\Delta
Q_{\ell}=(1/\pi)(\mathrm{d}\delta_{\ell}/\mathrm{d}k)$. The FSR embodies
the condition of overall charge neutrality, expected for a metallic
environment, as a consequence of the screening by all the electron
states within the Fermi sphere. In general form the rule may be
expressed as \cite{fri52,fri54}
\begin{equation}
Z_{1} =
\frac{2}{\pi} \, \sum_{\ell=0}^{\infty} (2\ell+1) \, \delta_{\ell}(k_{\mathrm{F}}),
\label{eq:FSR_exact}
\end{equation}
where $k_{\mathrm{F}}=(9\pi/4)^{1/3}(a_{0}r_{\mathrm{s}})^{-1}$ is the
Fermi wave vector. Here $a_{0}$ is the Bohr radius, and the
(dimensionless) one-electron radius (or Wigner--Seitz density parameter)
$r_{\mathrm{s}}$ is defined through the relation
$\frac{4}{3}\pi(r_{\mathrm{s}}a_{0})^{3}=n_{\mathrm{e}}^{-1}$. In this
circumstance of static screening the integral over the electron wave
vectors $k$ extends over a Fermi sphere of radius $k_{\mathrm{F}}$ (cf.\
\cite{zar95,lif98}). Without loss of generality we have not included
explicitly in Eq.~(\ref{eq:FSR_exact}) the contribution of the bound
electrons in which case the charge number $Z_{1}$ should be simply
replaced by $Z_{1}-N_{\mathrm{b}}$, being $N_{\mathrm{b}}$ the number of
bound electrons \cite{fri52,fri54}.

Within the B2 approximation the FSR (\ref{eq:FSR_exact}) reads
\begin{equation}
Z_{1} =
\frac{2}{\pi} \, \sum_{\ell=0}^{\infty} (2\ell+1) \,
\big[
\delta_{\ell,\mathrm{B}1}(k_{\mathrm{F}})
+ \delta_{\ell,\mathrm{B}2}(k_{\mathrm{F}})
\big],
\label{eq:FSR_B2_a}
\end{equation}
The right-hand side of the relation~(\ref{eq:FSR_B2_a}) can be evaluated
by substituting Eqs.~(\ref{eq:10}) and (\ref{eq:12}) for the first- and
second-order phase shifts, respectively. However, the simplest way is to
express the right-hand side of Eq.~(\ref{eq:FSR_B2_a}) through the
scattering amplitudes $f_{\mathrm{B}1}(k,0)$ and $f_{\mathrm{B}2}(k,0)$
for forward scattering (i.e., at $\theta=0$). From
Eqs.~(\ref{eq:f_B1_a}) and (\ref{eq:f_B2_a}) one obtains
\begin{equation}
\sum_{\ell=0}^{\infty} (2\ell+1)
\left[ \delta_{\ell,\mathrm{B}1}(k) + \delta_{\ell,\mathrm{B}2}(k) \right] =
k \, \Big\{ f_{\mathrm{B}1}(k,0) + \Rea\big[f_{\mathrm{B}2}(k,0)\big] \Big\}.
\label{eq:19}
\end{equation}
The forward-scattering amplitudes $f_{\mathrm{B}1}(k,0)$ and
$f_{\mathrm{B}2}(k,0)$ are easily evaluated from Eqs.~(\ref{eq:f_B1_c})
and (\ref{eq:f_B2_c}),
\begin{eqnarray}
f_{\mathrm{B}1}(k,0)
& = &
-\frac{m_{\mathrm{e}}}{2\pi\hbar^{2}} \widetilde{V}(0),
\label{eq:f0_B1}
\\
f_{\mathrm{B}2}(k,0)
& = &
\left( \frac{m_{\mathrm{e}}}{2\pi\hbar^{2}} \right)^{2}
\int\! \int
\frac{\mathrm{e}^{\mathrm{i}k\left\vert\mathbf{r}-\mathbf{r}'\right\vert}}
     {\left\vert\mathbf{r}-\mathbf{r}'\right\vert} \,
V(r) \, V(r') \,
\mathrm{e}^{\mathrm{i}k\hat{\mathbf{n}}\cdot(\mathbf{r}'-\mathbf{r})} \,
\mathrm{d}\mathbf{r} \, \mathrm{d}\mathbf{r}'.
\label{eq:f0_B2}
\end{eqnarray}
In Eq.~(\ref{eq:f0_B2}) we have made
$\mathbf{k}_{f}=\mathbf{k}_{i}=k\hat{\mathbf{n}}$, being
$\hat{\mathbf{n}}$ a unit vector, and $\widetilde{V}(0)$ is the Fourier
transform of the interaction potential at $q=0$. Thus, inserting
Eqs.~(\ref{eq:f0_B1}) and (\ref{eq:f0_B2}) into~(\ref{eq:19}) the FSR
(\ref{eq:FSR_B2_a}) within up to the B2 approximation is written as
\begin{eqnarray}
Z_{1}
& = &
\frac{2}{\pi} \, k_{\mathrm{F}} \,
\Big\{
f_{\mathrm{B}1}(k_{\mathrm{F}},0)
+ \Rea\big[f_{\mathrm{B}2}(k_{\mathrm{F}},0)\big]
\Big\}
\label{eq:FSR_B2_b}
\\
& = &
-\frac{m_{\mathrm{e}}k_{\mathrm{F}}}{\pi^{2}\hbar^{2}}
\left[
\widetilde{V}(0)
+ \frac{8\pi m_{\mathrm{e}}k_{\mathrm{F}}}{\hbar^{2}} \, \Sigma(k_{\mathrm{F}})
\right]
\label{eq:FSR_B2_c}
\end{eqnarray}
with
\begin{equation}
\Sigma(k) =
-\frac{1}{4(2\pi)^{2}k} \, \Rea
\left[
\int \! \int
\frac{\mathrm{e}^{\mathrm{i}kr'}}{r'} \,
V(r) \, V(|\mathbf{r}'+\mathbf{r}|) \,
\mathrm{e}^{\mathrm{i}k\hat{\mathbf{n}}\cdot\mathbf{r}'} \,
\mathrm{d}\mathbf{r} \, \mathrm{d}\mathbf{r}'
\right].
\label{eq:Sigma_a}
\end{equation}
In the last step of deriving Eq.~(\ref{eq:Sigma_a}) we have changed the
integration variable $\mathbf{r}'$ according to $\mathbf{r}'\to
\mathbf{r}'+\mathbf{r}$. Using now the Fourier transform
$\widetilde{V}(\mathbf{q})$ of the interaction potential, the function
$\Sigma(k)$ may be represented as
\begin{equation}
\Sigma(k) =
-\frac{1}{2(2\pi)^{4}k} \,
\int
\frac{|\widetilde{V}(\boldsymbol{\kappa})|^{2}\,\mathrm{d}\boldsymbol{\kappa}}
     {(\boldsymbol{\kappa}+\mathbf{k})^{2}-k^{2}},
\label{eq:Sigma_b}
\end{equation}
where we have used the property
$\widetilde{V}(-\mathbf{q})=\widetilde{V}^{\ast}(\mathbf{q})$. For a
spherically-symmetric interaction potential the Fourier transform is
also spherically symmetric,
$\widetilde{V}(\mathbf{q})=\widetilde{V}(q)$, and in addition it is a
real quantity. Then, after performing the angular integration in
Eq.~(\ref{eq:Sigma_b}) we finally arrive at
\begin{equation}
\Sigma(k) =
-\frac{1}{4(2\pi)^{3}k^{2}} \,
\int_{0}^{\infty} \widetilde{V}^{2}(\kappa) \,
\ln \left\vert \frac{\kappa+2k}{\kappa-2k} \right\vert \, \kappa \,
\mathrm{d}\kappa.
\label{eq:Sigma_c}
\end{equation}

Equation~(\ref{eq:FSR_B2_c}) is the main result of the present article.
The first and second terms in this expression are linear ($\sim Z_{1}$)
and quadratic ($\sim Z_{1}^{2}$) with respect to the interaction
potential representing, respectively, the first- and second-order Born
contributions to the FSR. While the first term in
Eq.~(\ref{eq:FSR_B2_c}) has been derived and studied previously
\cite{ech89,nag96,lif98} the second one is a new result that may be
viewed as a counterpart of the Barkas correction to the FSR. In general,
Eq.~(\ref{eq:FSR_B2_c}) is a transcendental equation that serves to
determine the screening length $\lambda\equiv 1/\alpha$, where $\alpha$
is the corresponding screening parameter, of the interaction potential
involved in $\widetilde{V}(0)$ and $\Sigma(k_{\mathrm{F}})$.

In what follows we apply the FSR in the B2 approximation,
Eq.~(\ref{eq:FSR_B2_c}), to the very important case of screened
potentials of the form
\begin{equation}
V(r) =
-\frac{Z_{1}e^{2}}{r} \, \Phi(\alpha r),
\label{eq:V_screened}
\end{equation}
where $\Phi(x)$ is the screening function. Then
\begin{equation}
\widetilde{V}(q) =
-(4\pi Z_{1}e^{2}/\alpha^{2}) \, \widetilde{\EuScript{V}}(q/\alpha),
\label{eq:28}
\end{equation}
where
\begin{equation}
\widetilde{\EuScript{V}}(x) =
\int_{0}^{\infty} \Phi(y) \, j_{0}(xy) \, y \, \mathrm{d}y
\label{eq:29}
\end{equation}
is the dimensionless Fourier transform of the interaction potential. In
terms of $\widetilde{\EuScript{V}}(x)$ we have
$\widetilde{V}(0)=-(4\pi Z_{1}e^{2}/\alpha^{2})\,\gamma$ with
\begin{equation}
\gamma =
\widetilde{\EuScript{V}}(0) =
\int_{0}^{\infty} \Phi(x) \, x \, \mathrm{d}x
\label{eq:gamma}
\end{equation}
and
$\Sigma(k_{\mathrm{F}})=-(Z^{2}_{1}e^{4}/4k_{\mathrm{F}}^{2}\alpha^{2})\,g(u)$
with $u=2k_{\mathrm{F}}/\alpha$ and
\begin{equation}
g(u) =
\frac{2}{\pi} \,
\int_{0}^{\infty} \widetilde{\EuScript{V}}^{2}(x) \,
\ln \left\vert \frac{x+u}{x-u} \right\vert \, x \, \mathrm{d}x.
\label{eq:g_a}
\end{equation}
Inserting $\widetilde{\EuScript{V}}(x)$, Eq.~(\ref{eq:29}),
into~(\ref{eq:g_a}) and using expression~(\ref{eq:A.4}) in~\ref{sec:app1}
we may, alternatively, write $g(u)$ in coordinate space via the relation
\begin{equation}
g(u) =
\int_{0}^{\infty} \Phi(x) \, \mathrm{d}x \,
\int_{0}^{\infty} \Big[ \si\!\big(u(x+y)\big) - \si\!\big(u|x-y|\big) \Big] \,
\Phi(y) \, \mathrm{d}y,
\label{eq:g_b}
\end{equation}
where $\si(x)$ is the sine integral \cite{abr72,gra80}. Introducing the
quantities $\widetilde{V}(0)$ and $\Sigma(k_{\mathrm{F}})$ defined above
for an arbitrary screened potential into Eq.~(\ref{eq:FSR_B2_c}) it
turns out that the general solution for the screening parameter $\alpha$
in the B2 approximation may be cast in the form
\begin{equation}
\alpha =
\alpha_{\mathrm{RPA}}
\left[ \gamma + \frac{\pi}{2}\,Z_{1}\chi^{2}\,g(u) \right]^{1/2},
\label{eq:alpha_B2}
\end{equation}
where
$\alpha_{\mathrm{RPA}}=1/\lambda_{\mathrm{TF}}=(4k_{\mathrm{F}}/\pi a_{0})^{1/2}$
is the screening parameter within the RPA ($\lambda_{\mathrm{TF}}$ is
the Thomas--Fermi screening length) and $\chi^{2}=(\pi
k_{\mathrm{F}}a_{0})^{-1}$ is the (dimensionless) Lindhard density
parameter of the DEG. The numerical constant $\gamma$ and the function
$g(u)$ should be specified for the adopted screened potential according
to Eqs.~(\ref{eq:gamma}) and (\ref{eq:g_a}) [or (\ref{eq:g_b})],
respectively.

In Eq.~(\ref{eq:alpha_B2}) the term containing the function $g(u)$ is
the second-order Born correction to the screening parameter $\alpha$.
Neglecting this term we retrieve the well-known expression
\begin{equation}
\alpha_{\mathrm{B}1} =
\alpha_{\mathrm{RPA}} \, \gamma^{1/2}
\label{eq:alpha_B1}
\end{equation}
for $\alpha$ derived within the B1 approximation (see, e.g.,
\cite{lif98}). It is noteworthy that, in contrast to the
first-order $\alpha_{\mathrm{B}1}$, the second-order screening parameter
(\ref{eq:alpha_B2}) depends on $Z_{1}$ and thus predicts different
screening parameters for attractive ($Z_{1}>0$) and repulsive
($Z_{1}<0$) electron-ion interactions.
It should also be stressed that Eq.~(\ref{eq:alpha_B1}) represents the
solution of Eq.~(\ref{eq:alpha_B2}) at high electron densities, i.e.,
when $\chi^{2}\ll 1$.

It is also useful to study the asymptotic solutions of
Eq.~(\ref{eq:alpha_B2}) at small electron densities, $\chi^{2}\gg 1$. In
order to investigate these solutions we realize that the function $g(u)$
in that equation behaves as $g(u)\simeq C_{0}u$ when $u\ll 1$ and as
$g(u)\simeq C_{\infty}/u$ when $u\gg 1$, where $C_{0}$ and $C_{\infty}$
are numerical constants. The validity of these asymptotic forms of
$g(u)$ becomes evident from the general Eqs.~(\ref{eq:g_a}) and
(\ref{eq:29}) at $u\ll 1$ and $u\gg 1$ for an arbitrary screening
function $\Phi(x)$
\begin{eqnarray}
C_{0}
& = &
\frac{4}{\pi} \,
\int_{0}^{\infty} \widetilde{\EuScript{V}}^{2}(x) \, \mathrm{d}x =
4\int_{0}^{\infty} \Phi(x) \, x \, \mathrm{d}x \,
\int_{x}^{\infty} \Phi(y) \, \mathrm{d}y,
\label{eq:C0}
\\
C_{\infty}
& = &
\frac{4}{\pi} \,
\int_{0}^{\infty} \widetilde{\EuScript{V}}^{2}(x) \, x^{2} \, \mathrm{d}x =
2\int_{0}^{\infty} \Phi^{2}(x) \, \mathrm{d}x.
\label{eq:Cinfinity}
\end{eqnarray}
Then, at small electron densities ($\chi^{2}\gg 1$) and for a positive
ion ($Z_{1}>0$) from Eq.~(\ref{eq:alpha_B2}) it follows that
\begin{equation}
\frac{\alpha}{\alpha_{\mathrm{RPA}}} \simeq
\left( \frac{\pi}{2}\,C_{0}\,Z_{1}\chi \right)^{1/3};
\label{eq:alpha_B2_asymptotic1}
\end{equation}
hence, the ratio $\alpha/\alpha_{\mathrm{RPA}}$ increases as $\sim
(Z_{1}\chi)^{1/3}$. In turn, at small densities ($\chi^{2}\gg 1$) but
for a negative ion ($Z_{1}<0$) assuming that
$\alpha/\alpha_{\mathrm{RPA}}<1/\chi$ from Eq.~(\ref{eq:alpha_B2}) we
get the asymptotic solution
\begin{equation}
\frac{\alpha}{\alpha_{\mathrm{RPA}}} \simeq
\frac{\gamma}{\frac{\pi}{2}\,C_{\infty}\,|Z_{1}|\chi^{3}}
\label{eq:alpha_B2_asymptotic2}
\end{equation}
so that $\alpha/\alpha_{\mathrm{RPA}}$ decreases as
$\sim (|Z_{1}|\chi^{3})^{-1}$.

Equation~(\ref{eq:alpha_B2}) determines the second-order screening
parameter as a function of the density of the DEG and the atomic number
of the ion. Recalling that the second-order correction [i.e., the second
term in Eq.~(\ref{eq:alpha_B2})] should be smaller than the first one,
Eq.~(\ref{eq:alpha_B2}) can be solved iteratively. A simple estimate is
achieved if $\alpha_{\mathrm{B}1}$, Eq.~(\ref{eq:alpha_B1}), is inserted
into the second-order term of Eq.~(\ref{eq:alpha_B2}), yielding
\begin{eqnarray}
\alpha
& \simeq &
\alpha_{\mathrm{RPA}}
\left[
\gamma + \frac{\pi}{2}\,Z_{1}\chi^{2}\,g(u_{\mathrm{B}1})
\right]^{1/2}
\label{eq:alpha_B2_approx1}
\\
&\simeq &
\alpha_{\mathrm{RPA}} \, \gamma^{1/2}
\left[ 1 + \frac{\pi}{4\gamma}\,Z_{1}\chi^{2}\,g(u_{\mathrm{B}1}) \right],
\label{eq:alpha_B2_approx2}
\end{eqnarray}
where
$u_{\mathrm{B}1}=2k_{\mathrm{F}}/\alpha_{\mathrm{B}1}=(\gamma\chi^{2})^{-1/2}$.
The second term in Eq.~(\ref{eq:alpha_B2_approx2}) is indeed small for
typical densities of conduction electrons in metals ($1.5\lesssim
r_{\mathrm{s}}\lesssim 5$) and for proton and antiproton projectiles
($Z_{1}=\pm 1$).

For practical applications we include below explicit examples of
interaction potentials which are widely used to model the stopping of
low-energy ions in a DEG with either quantum or classical formalisms,
namely the Yukawa, hydrogenic, Hulth\'{e}n, and Mensing potentials. For
instance, the first three of them have been adopted by several authors
to describe the slowing down of positive ions in
solids~\cite{mei75,fer77,apa87,ven88,sor90,cal93,cal94,nag99,vin08},
whereas the latter has proven suitable to model the stopping power of
antiprotons~\cite{nag94,ari04,ner05}. We derive the respective numerical
constants $\gamma$ and functions $g(u)$ in
Sections~\ref{sec:yuk}--\ref{sec:men} (and in~\ref{sec:app1}). For the sake
of completeness we summarize them in Table~\ref{tab:1} together with the
numerical constants $C_{0}$ and $C_{\infty}$. The evaluation of these
constants is trivial except in the case of the Hulth\'{e}n potential,
see~\ref{sec:app2}.

\begin{table*}
\caption{The parameter $\gamma$ and the expressions for function $g(u)$ as
well as the numerical constants $C_{0}$ and $C_{\infty}$ determining the
asymptotic behaviors of the function $g(u)$ at $u\ll 1$ and $u\gg 1$,
respectively, for the indicated potentials.}
\label{tab:1}
\begin{center}
\begin{tabular}{lllll}
\hline
$V(r)$      & $\gamma$    & $g(u)$
            & $C_{0}$     & $C_{\infty}$                       \\
\hline
Yukawa      & $1$         & Eq.~(\ref{eq:g_Yukawa})
            & $1$         & $1$                                \\
hydrogenic  & $2$         & Eq.~(\ref{eq:g_hydrogenic})
            & $25/8$      & $13/8$                             \\
Hulth\'{e}n & $2\zeta(3)$ & Eqs.~(\ref{eq:g_b}) and (\ref{eq:Phi_Hulthen})
            & $4\zeta(5)$ & $4[\pi^{2}/6-\zeta(3)]$            \\
Mensing     & $1/6$       & Eq.~(\ref{eq:g_Mensing})
            & $1/10$      & $2/3$                              \\
\hline
\end{tabular}
\end{center}
\end{table*}


\subsection{Yukawa potential}
\label{sec:yuk}

As a first example we choose the Yukawa interaction potential, whose
screening function is given by
\begin{equation}
\Phi(x) = \mathrm{e}^{-x}.
\label{eq:Phi_Yukawa}
\end{equation}
The Fourier transform of this potential is
\begin{equation}
\widetilde{V}(q) =
-\frac{4\pi Z_{1}e^{2}}{q^{2}+\alpha^{2}}.
\label{eq:V_Fourier_Yukawa}
\end{equation}
Substituting the screening function $\Phi(x)$ and the Fourier transform
(\ref{eq:V_Fourier_Yukawa}) into Eqs.~(\ref{eq:gamma}) and
(\ref{eq:g_a}), respectively, we arrive at $\gamma=1$ \cite{ech89,lif98}
and
\begin{equation}
g(u) =
\frac{2}{\pi} \,
\int_{0}^{\infty} \ln \left\vert \frac{z+u}{z-u} \right\vert \,
\frac{z\,\mathrm{d}z}{\left(z^{2}+1\right)^{2}}.
\label{eq:43}
\end{equation}
Equation~(\ref{eq:43}) is further simplified integrating by parts, which
finally yields
\begin{equation}
g(u) =
\frac{u}{u^{2}+1}.
\label{eq:g_Yukawa}
\end{equation}


\subsection{Hydrogenic potential}
\label{sec:hyd}

The screening function of the hydrogenic potential \cite{apa87} is
\begin{equation}
\Phi(x) = \left( 1 + \frac{1}{2}\,x \right) \mathrm{e}^{-x}.
\label{eq:Phi_hydrogenic}
\end{equation}
The Fourier transform of the hydrogenic potential reads
\begin{equation}
\widetilde{V}(q) =
-4\pi Z_{1}e^{2} \,
\frac{q^{2}+2\alpha^{2}}{\left(q^{2}+\alpha^{2}\right)^{2}}.
\label{eq:V_Fourier_hydrogenic}
\end{equation}
In this case $\gamma=2$ \cite{ech89,lif98} and
\begin{equation}
g(u) =
p^{2} \left[ T_{2}(p) + 2p^{2}\,T_{3}(p) + p^{4}\,T_{4}(p) \right],
\label{eq:47}
\end{equation}
where $p=\alpha/2k_{\mathrm{F}}=1/u$, and
\begin{equation}
T_{n}(p) =
\frac{2}{\pi} \,
\int_{0}^{\infty} \ln \left\vert \frac{z+1}{z-1} \right\vert \,
\frac{z\,\mathrm{d}z}{\left(z^{2}+p^{2}\right)^{n}}
\label{eq:48}
\end{equation}
with $n=$ 2, 3, and 4. $T_{2}(p)=(1/p^{2})f_{0}(1/p)$ with $f_{0}(u)$
given by Eq.~(\ref{eq:g_Yukawa}), whereas $T_{3}(p)$ and $T_{4}(p)$ can
be evaluated from the relations
$T_{3}(p)=-\frac{1}{4p}T_{2}^{\prime}(p)$ and
$T_{4}(p)=-\frac{1}{6p}T_{3}^{\prime}(p)$, respectively. Finally,
substituting the functions $T_{n}(p)$ into Eq.~(\ref{eq:47}) one gets
\begin{equation}
g(u) =
\frac{u}{u^{2}+1}
\left[ \frac{13}{8} + \frac{7}{6\left(
u^{2}+1\right) }+\frac{1}{3\left( u^{2}+1\right)^{2}} \right].
\label{eq:g_hydrogenic}
\end{equation}


\subsection{Hulth\'{e}n potential}
\label{sec:hul}

The screening function pertaining to the Hulth\'{e}n potential
\cite{hul42A,hul42B} (see also \cite{gla12}) is
\begin{equation}
\Phi(x) = x \, (\mathrm{e}^{x}-1)^{-1}.
\label{eq:Phi_Hulthen}
\end{equation}
From Eqs.~(\ref{eq:gamma}) and (\ref{eq:Phi_Hulthen}) one finds
$\gamma=2\zeta(3)$ \cite{ech89}, where $\zeta(z)$ is the Riemann zeta
function \cite{abr72}. On the other hand, $g(u)$ must be computed
numerically by means of Eqs.~(\ref{eq:g_b}) and (\ref{eq:Phi_Hulthen}).


\subsection{Mensing potential}
\label{sec:men}

For the Mensing potential \cite{men27}
\begin{equation}
\Phi(x) = (1-x)\,\Theta(1-x),
\label{eq:Phi_Mensing}
\end{equation}
where $\Theta(x)$ is the Heaviside unit-step function. Then
\begin{equation}
\widetilde{V}(q) =
-\frac{4\pi Z_{1}e^{2}}{q^{2}} \, \big[ 1 - j_{0}(q/\alpha) \big].
\label{eq:V_Fourier_Mensing}
\end{equation}
Now $\gamma=1/6$ and (see~\ref{sec:app1} for details)
\begin{eqnarray}
g(u)
& = &
\left( \frac{1}{2u^{2}} - \frac{1}{3} \right) \frac{1-j_{0}(2u)}{u}
\nonumber
\\
& &
\mbox{}
+ \frac{2}{3} \left[ \left( \frac{1}{2u} - u \right)
j_{0}^{2}(u) - \left( \frac{1}{u} - \frac{u}{2} \right)
j_{0}^{2}(u/2) \right]
\nonumber
\\
& &
\mbox{}
+ \frac{2}{3} \left[ \frac{1-j_{0}(u)}{u} + \si(2u) - \si(u) \right].
\label{eq:g_Mensing}
\end{eqnarray}


\section{Comparison with exact numerical solutions}
\label{sec:4}

We present now results for the Yukawa, hydrogenic, Hulth\'{e}n, and
Mensing potentials using the theoretical findings of Sections~\ref{sec:2}
and \ref{sec:3}. Bare ions with charges $Z_{1}=\pm 1$, $\pm 2$, and $\pm 3$
shall be considered along with a wide range of electron densities,
$r_{\mathrm{s}}\leqslant 5$, with the one-electron radius connected to
the Lindhard density parameter through the relation
$r_{\mathrm{s}}=(9\pi^{4}/4)^{1/3}\chi^{2}$. Some values of $Z_{1}$ may
be unrealistic for certain interaction potentials and are analyzed here
just to test the B2 approximation both for negative and positive
$Z_{1}$.

Exact screening parameters have also been computed for the
aforementioned combinations of $V(r)$, $Z_{1}$, and $r_{\mathrm{s}}$. To
do so, phase shifts up to $\ell=100$ were evaluated by solving
numerically the Calogero equation (\ref{eq:Calogero}). Then, a
self-consistent iterative procedure adjusted the value of $\alpha$ so
that the ensuing absolute phase shifts satisfy the exact FSR,
Eq.~(\ref{eq:FSR_exact}).

\begin{figure*}[htbp]
\begin{center}
\includegraphics[width=0.95\textwidth]{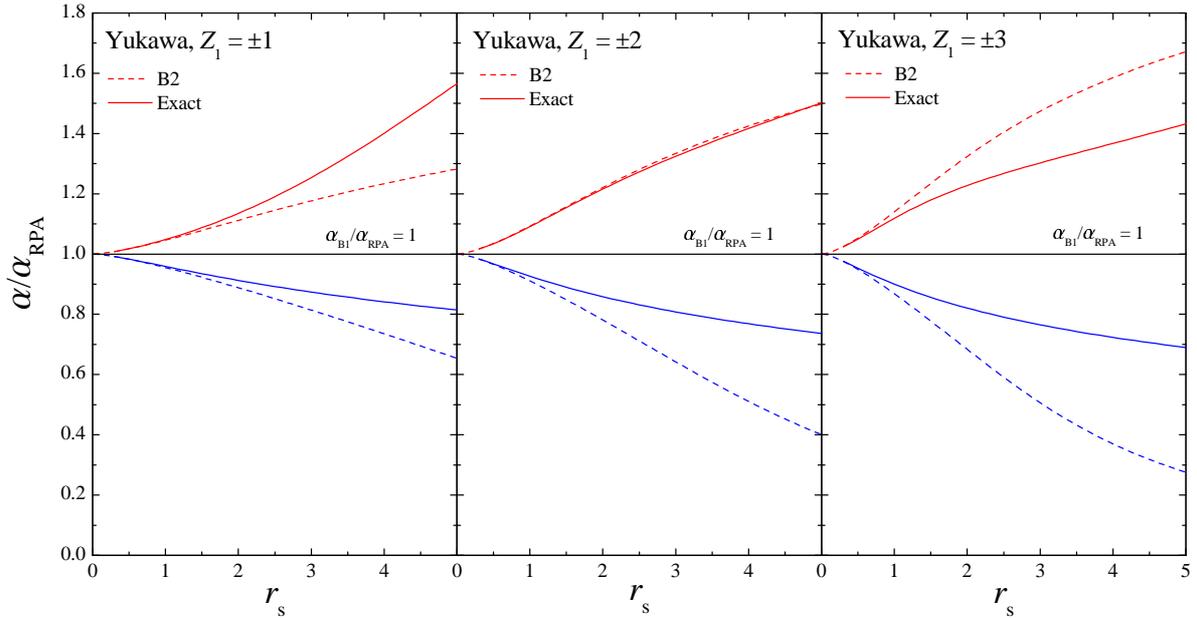}
\end{center}
\caption{Normalized screening parameter $\alpha/\alpha_{\mathrm{RPA}}$
as a function of $r_{\mathrm{s}}$ for the Yukawa potential and for
$Z_{1}=\pm 1,\,\pm 2,\,\pm 3$. Shown are the solutions of
Eq.~(\ref{eq:alpha_B2}) within the B2 approximation (dashed curves) and
the exact FSR~(\ref{eq:FSR_exact}) (solid curves). The horizontal solid
line indicates $\alpha_{\mathrm{B}1}/\alpha_{\mathrm{RPA}}=\gamma^{1/2}$
which clearly delimits the domains with positive (upper region) and
negative (lower region) values of $Z_{1}$.}
\label{fig:1}
\end{figure*}

\begin{figure*}[htbp]
\begin{center}
\includegraphics[width=0.95\textwidth]{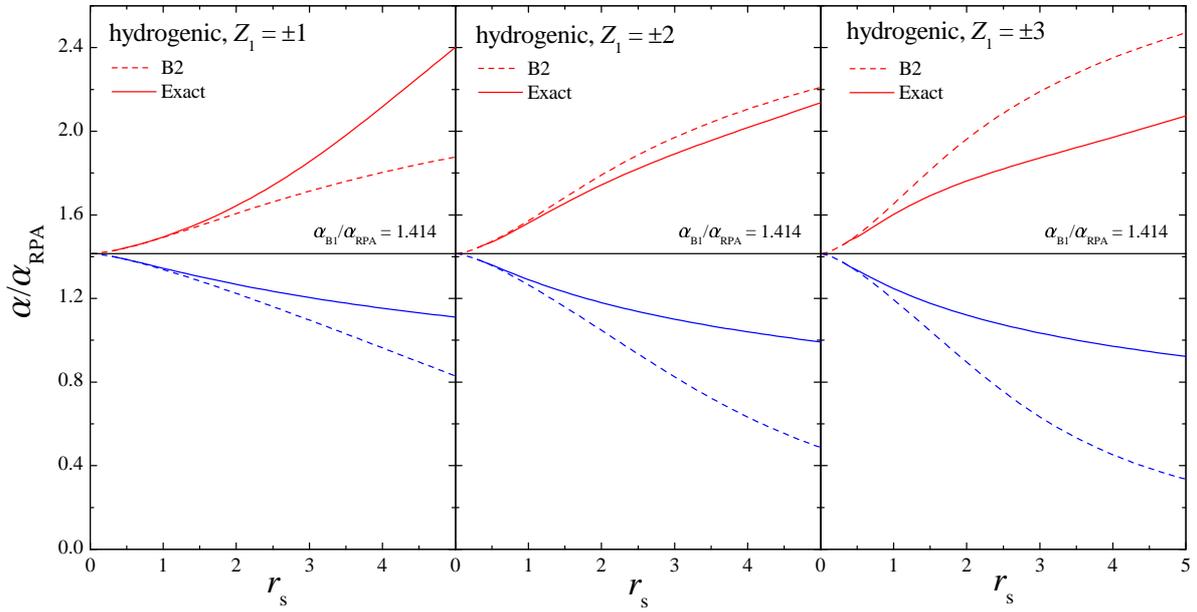}
\end{center}
\caption{Same as in Fig.~\ref{fig:1} but for the hydrogenic potential.}
\label{fig:2}
\end{figure*}

\begin{figure*}[htbp]
\begin{center}
\includegraphics[width=0.95\textwidth]{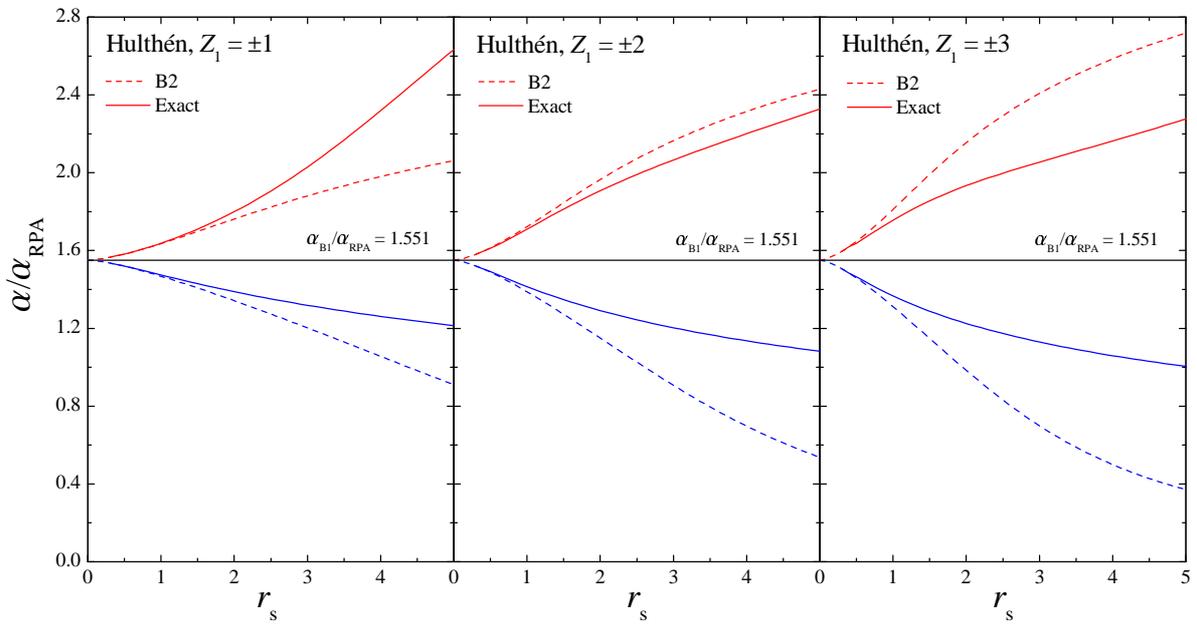}
\end{center}
\caption{Same as in Fig.~\ref{fig:1} but for the Hulth\'{e}n potential.}
\label{fig:3}
\end{figure*}

\begin{figure*}[htbp]
\begin{center}
\includegraphics[width=0.95\textwidth]{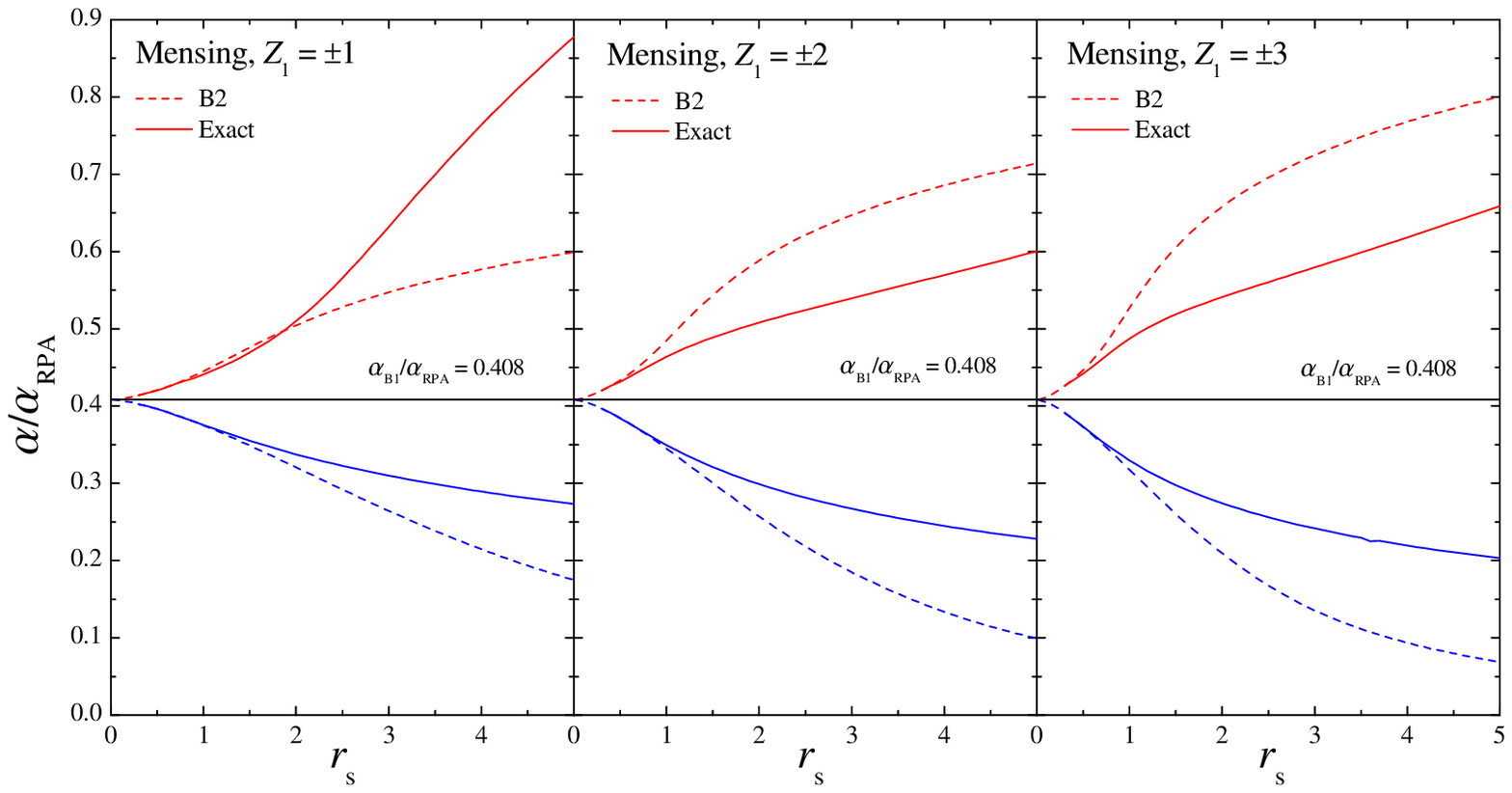}
\end{center}
\caption{Same as in Fig.~\ref{fig:1} but for the Mensing potential.}
\label{fig:4}
\end{figure*}

Figs.~\ref{fig:1}--\ref{fig:4} display, for the investigated
interaction potentials, the ratios $\alpha/\alpha_{\mathrm{RPA}}$ as a
function of $r_{\mathrm{s}}$ and $Z_{1}$. It should be emphasized that
$\alpha_{\mathrm{RPA}}$ does not depend on $Z_{1}$ but varies with
$r_{\mathrm{s}}$ ($\alpha_{\mathrm{RPA}}\propto r_{\mathrm{s}}^{-1/2}$).
The plotted data are the predictions of the B2 approximation (dashed
curves), given by Eq.~(\ref{eq:alpha_B2}), and the exact screening
parameters (solid curves). The $\alpha/\alpha_{\mathrm{RPA}}$ curves
belonging to positive and negative ions are separated by the horizontal
lines $\alpha_{\mathrm{B}1}/\alpha_{\mathrm{RPA}}=\gamma^{1/2}$ (see
Table~\ref{tab:1} for the specific values of $\gamma$). It is noteworthy
that $\gamma^{1/2}$ varies significantly for the studied potentials. The
smallest and the largest values of $\gamma^{1/2}$ occur for the Mensing
and Hulth\'{e}n potentials, respectively, with the latter being almost
4~times larger than the former. On the other hand, the B2 approximation
does introduce a dependence of $\alpha$ on $Z_{1}$. In fact,
Eq.~(\ref{eq:alpha_B2}) correctly predicts that
$\alpha>\alpha_{\mathrm{B}1}$ if $Z_{1}>0$ and
$\alpha<\alpha_{\mathrm{B}1}$ if $Z_{1}<0$.

From Figs.~\ref{fig:1}--\ref{fig:4} it is seen that the B2 screening
parameters are in excellent agreement with the exact values when
$r_{\mathrm{s}}\lesssim 1$, i.e., at large and moderate electron
densities. Moreover, in the extreme regimes with $r_{\mathrm{s}}\to 0$,
which are of interest for degenerate astrophysical plasmas, the B1
approximation is increasingly accurate and in this limit both treatments
yield $\alpha/\alpha_{\mathrm{RPA}}=\gamma^{1/2}$. In the opposite case
of lower densities, $r_{\mathrm{s}}\gtrsim 1$, the B2 approximation
deviates from the self-consistent results of the exact FSR but it
improves upon the $Z_{1}$-independent $\alpha_{\mathrm{B}1}$. Besides,
the B2 approximation underestimates systematically the screening
parameter when $Z_{1}<0$ (see Figs.~\ref{fig:1}--\ref{fig:4}). In the
case of positive ions, however, this approximation generally
underestimates $\alpha$ for $Z_{1}= +1$ while overestimating it for
$Z_{1}=+2$ and $+3$. Interestingly, the B2 approximation for the Yukawa
potential with $Z_{1}=+2$ almost coincides with the exact results
(Fig.~\ref{fig:1}), albeit this is an accidental agreement and is not
observed for other potentials. In~\ref{sec:app3} we further discuss the
B2 approximation for the screening parameter and explore a simple
improvement of the model based on the $[1/1]$ Pad\'{e} approximant.


\section{Conclusions}
\label{sec:5}

We have proposed a simple way to calculate the static screening
parameter (the inverse of the screening length) for an ion in a DEG
based on the B2 approximation for the FSR and on the use of this sum
rule to adjust in a self-consistent manner the screening parameters of
the various interaction potentials. The developed model furnishes a
simple scheme to incorporate the effects of the non-linear ion-solid
coupling in the quantum formulation of screening and scattering
processes, which is regarded an appropriate framework to describe
non-linear screening and energy loss of ions in solids.

In the high-density limit ($r_{\mathrm{s}}\lesssim 1$) the present
approach agrees excellently with the exact screening parameters
calculated self-consistently by imposing the FSR requirement to the
numerical phase shifts. At intermediate and small densities
($r_{\mathrm{s}}\gtrsim 1$) our results depart progressively from the
exact values but still improve upon the $Z_{1}$-independent predictions
of the B1 approximation. More precisely, at $r_{\mathrm{s}}\gtrsim 1$
our model systematically underestimates the screening parameters for
negative ions as well as for $Z_{1}=+1$ compared to the exact treatment,
while overestimating them for $Z_{1}=+2$ and $+3$.

The Pad\'{e} approximant to the Born series in the perturbative FSR has
been addressed in~\ref{sec:app3} as the simplest way to improve the
present second-order Born approximation. It is found that the Pad\'{e}
approximant of order $[1/1]$ systematically shifts the perturbative
screening parameters towards higher values thus yielding better
predictions for any negative ion as well as for $Z_{1}=+1$.
Nevertheless, it impairs the agreement with the exact theory for ions
with $Z_{1}=+2$ and $+3$.

The model also provides the possibility to calculate the screening
parameter of a static impurity ion immersed in a two-dimensional (2D)
electron gas. In this case the starting point should be the 2D
Schr\"{o}dinger equation accompanied by an appropriate FSR adapted to
the 2D geometry \cite{ste67}. Moreover, bearing in mind some practical
applications the present perturbative formalism can be extended easily
to account for the dynamical screening effects of swift ions in solids
as proposed in~\cite{lif98}. An expected consequence of the dynamical
effects is the replacement of the parameter $\gamma$ and the function
$g(u)$ in Eq.~(\ref{eq:alpha_B2}) by velocity-dependent ones. The
validity of the resulting dynamical and perturbative model will be
limited by the restriction in the interaction potential, which is
usually assumed to maintain the spherical symmetry for a moving ion.
However, the self-consistent adjustment of the potential makes this
assumption less critical, as may be checked by considering the behavior
of the stopping power of the ion in the more unfavorable case of high
velocities~\cite{lif98}.


\section*{Acknowledgements}

The work of H.B.~Nersisyan has been supported by the State Committee of
Science of the Armenian Ministry of Higher Education and Science (Project No.\ 13-1C200).
J.M.~Fern\'{a}ndez-Varea thanks the financial support from the Spanish
Ministerio de Ciencia e Innovaci\'{o}n (Project No.\
FPA2009-14091-C02-01) and FEDER.


\appendix

\section{Evaluation of $g(u)$ for the Mensing potential}
\label{sec:app1}

In this Appendix we evaluate the function $g(u)$ which determines the
second-order correction in the screening parameter (\ref{eq:alpha_B2})
for the Mensing potential. Inserting Eq.~(\ref{eq:V_Fourier_Mensing})
into Eq.~(\ref{eq:g_a}) we have
\begin{equation}
g(u) = -\frac{2}{\pi}
\int_{0}^{\infty} \ln \left\vert \frac{z+u}{z-u}\right\vert \,
G^{\prime}(z) \, \mathrm{d}z,
\label{eq:A.1}
\end{equation}
where
\begin{eqnarray}
G(z)
& = &
\int_{z}^{\infty} \left[ 1 - j_{0}(x) \right]^{2} \frac{\mathrm{d}x}{x^{3}}
\nonumber
\\
& = &
\frac{1}{12z^{2}} \left[ 2j_{0}(2z)-8j_{0}(z)+3+3j_{0}^{2}(z) \right]
\nonumber
\\
& &
\mbox{}
+ \frac{1}{3} \left\{ \frac{1}{2} \left[ j_{0}^{2}(z/2) - j_{0}^{2}(z) \right]
+ j_{0}(z) - j_{0}(2z) + \ci(2z) - \ci(z) \right\}.
\label{eq:A.2}
\end{eqnarray}
Here $\ci(z)$ is the cosine integral \cite{abr72,gra80}. The explicit
form for the function $G(z)$ in Eq.~(\ref{eq:A.2}) is obtained by
standard integration techniques \cite{abr72,gra80}. Next we integrate
Eq.~(\ref{eq:A.1}) by parts and get
\begin{equation}
g(u) =
\frac{4u}{\pi} \, \int_{0}^{\infty} \frac{G(z)}{u^{2}-z^{2}} \, \mathrm{d}z.
\label{eq:A.3}
\end{equation}
The singularity at $z=u$ in Eq.~(\ref{eq:A.3}) must be understood in the
sense of Cauchy's principal value. Further progress is achieved by
employing the following integral for $\ci(az)$ \cite{gra80}
\begin{equation}
\frac{2}{\pi} \, \int_{0}^{\infty} \frac{\ci(az)}{u^{2}-z^{2}} \, \mathrm{d}z =
\frac{1}{u} \, \si(au).
\label{eq:A.4}
\end{equation}
Similar integrals can be found in~\cite{gra80} for $j_{0}(az)$ as
well as for $j_{0}^{2}(az)$. Then the contribution of the first term (in
the square brackets) of Eq.~(\ref{eq:A.2}) to (\ref{eq:A.3}) becomes
\begin{equation}
g_{\mathrm{I}}(u) =
\frac{C}{u} + \frac{1}{3\pi u} \, \int_{0}^{\infty}
\left[ 2j_{0}(2z) - 8j_{0}(z) + 3j_{0}^{2}(z) \right]
\frac{\mathrm{d}z}{u^{2}-z^{2}},
\label{eq:A.5}
\end{equation}
where
\begin{equation}
C =
\frac{1}{3\pi} \, \int_{0}^{\infty} \left[
2j_{0}(2z)-8j_{0}(z)+3+3j_{0}^{2}(z)\right] \frac{\mathrm{d}z}{z^{2}}
\label{eq:A.6}
\end{equation}
is a numerical constant which can be represented in an alternative
manner by grouping the different terms in Eq.~(\ref{eq:A.6}),
\begin{equation}
C =
\frac{1}{3\pi} \left\{
4\int_{0}^{\infty} \left[ 1 - j_{0}(z) \right]
\frac{\mathrm{d}z}{z^{2}}
- 3 \int_{0}^{\infty} \left[ 1 - j_{0}^{2}(z) \right]
\frac{\mathrm{d}z}{z^{2}} \right\}.
\label{eq:A.7}
\end{equation}
In Eq.~(\ref{eq:A.7}) the first and the second integrals are equal to
$\pi/4$ and $\pi/3$, respectively \cite{gra80}. Thus $C=0$ and the
function $g_{\mathrm{I}}(u)$ in Eq.~(\ref{eq:A.5}) is determined by the
second term only which, using the known integrals for the spherical
Bessel functions \cite{gra80}, is evaluated in the explicit form
\begin{equation}
g_{\mathrm{I}}(u) =
\frac{1}{2u^{3}} \, \big[ 1-j_{0}(2u) \big]
+ \frac{1}{3u} \, \big[ j_{0}^{2}(u) - 2j_{0}^{2}(u/2) \big].
\label{eq:A.8}
\end{equation}
The contributions of the other terms of $G(z)$ to $g(u)$ are calculated
analogously using Eq.~(\ref{eq:A.4}) and similar integrals for the
spherical Bessel functions. The final result is Eq.~(\ref{eq:g_Mensing}).


\section{Evaluation of $C_{0}$ and $C_{\infty}$ for the Hulth\'{e}n potential}
\label{sec:app2}

In the case of the Hulth\'{e}n
potential the evaluation of the constants $C_{0}$ and $C_{\infty}$ is
performed using the second equalities of Eqs.~(\ref{eq:C0}) and
(\ref{eq:Cinfinity}) with (\ref{eq:Phi_Hulthen}). For $C_{0}$ this
yields
\begin{equation}
C_{0} =
8 \left[ h_{3}^{(2)} + 3h_{4}^{(1)} - 4\zeta(5) \right],
\label{eq:B.1}
\end{equation}
where $h_{r}^{(s)}=\sum_{n=1}^{\infty}n^{-r}H_{n}^{(s)}$ (with positive
integers $r\geqslant 2$ and $s$), $H_{n}^{(s)}$ are the so-called
harmonic numbers, $H_{n}^{(s)}=\sum_{k=1}^{n}k^{-s}$, and $\zeta(x)$ is
the Riemann zeta function. Equation~(\ref{eq:B.1}) is deduced by making
a power series expansion of both functions $\Phi (x)$ and $\Phi (y)$ for
the Hulth\'{e}n potential with respect to $\mathrm{e}^{-x}$ and $\mathrm{e}^{-y}$,
respectively [see Eq.~(\ref{eq:Phi_Hulthen})]. Inserting these series
into Eq.~(\ref{eq:C0}) and after a few algebraic manipulations we arrive
at Eq.~(\ref{eq:B.1}). The remaining steps in the derivation of $C_{0}$
are straightforward. First, using an obvious property of the harmonic
numbers, $H_{n}^{(s)}=H_{n-1}^{(s)}+n^{-s}$, we see that
\begin{equation}
h_{r}^{(s)} =
\zeta(r+s) + \sum_{n=2}^{\infty} \frac{1}{n^{r}}\,H_{n-1}^{(s)}.
\label{eq:B.2}
\end{equation}
Next, applying repeatedly this relation a system of algebraic equations
for the three quantities, $h_{2}^{(3)}$, $h_{3}^{(2)}$, and
$h_{4}^{(1)}$ can be set up. The solution of this system gives
$h_{3}^{(2)}=3\zeta(2)\zeta(3)-(9/2)\zeta(5)$ and
$h_{4}^{(1)}=3\zeta(5)-\zeta(2)\zeta(3)$. Inserting these quantities
into Eq.~(\ref{eq:B.1}) we finally arrive at $C_{0}=4\zeta(5)$.

The evaluation of the constant $C_{\infty}$ is facilitated by using the
known integral (see, e.g., \cite{gra80})
\begin{equation}
\Im_{n}(a) =
\int_{0}^{\infty} \frac{x^{n}\,\mathrm{d}x}{\mathrm{e}^{ax}-1} =
\frac{n!}{a^{n+1}} \, \zeta(n+1),
\label{eq:B.3}
\end{equation}
where $n$ is an arbitrary positive integer and $a>0$. Then $C_{\infty}$
is expressed via $\Im_{1}(a)$ and $\Im_{2}(a)$ as follows
\begin{equation}
C_{\infty} =
-2 \left[ \frac{\mathrm{d}\Im_{1}(a)}{\mathrm{d}a} + \Im_{2}(a) \right]_{a=1} =
4 \left[ \frac{\pi^{2}}{6} - \zeta(3) \right].
\label{eq:B.4}
\end{equation}


\section{Pad\'{e} approximant}
\label{sec:app3}

\begin{figure*}[htbp]
\begin{center}
\includegraphics[width=0.95\textwidth]{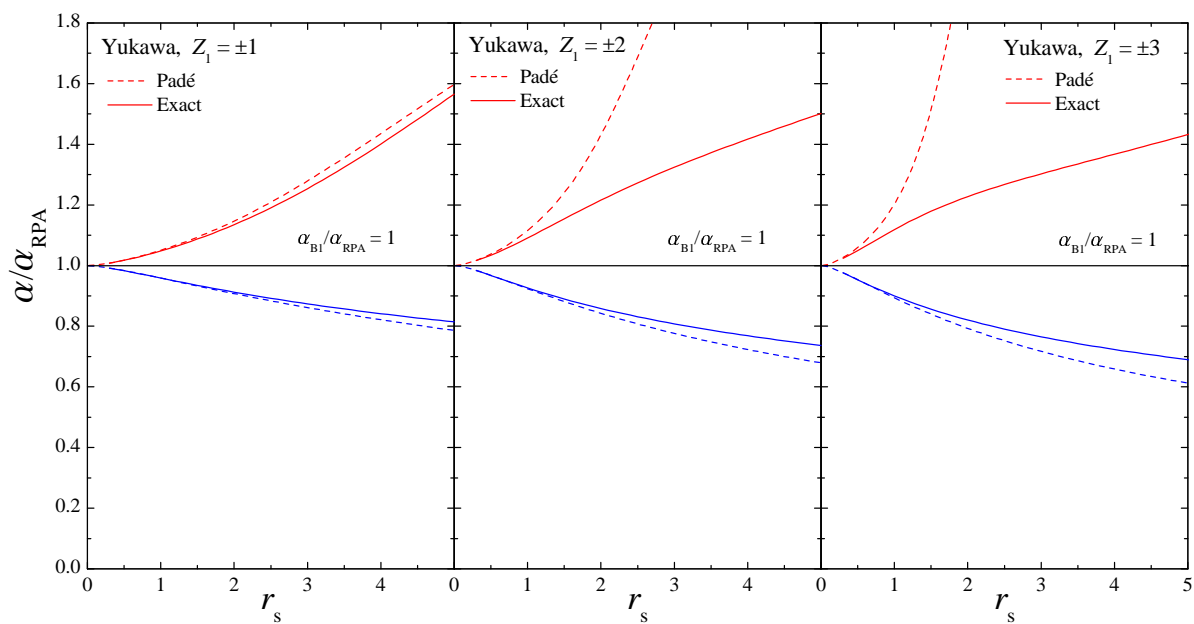}
\end{center}
\caption{The normalized screening parameter
$\alpha/\alpha_{\mathrm{RPA}}$ as a function of the density parameter
$r_{\mathrm{s}}$ for the Yukawa potential and for $Z_{1}=\pm 1,\;\pm
2,\;\pm 3$. Shown are the solutions of Eq.~(\ref{eq:pade1}) (i.e., the
Pad\'{e} approximant of Eq.~(\ref{eq:FSR_B2_c})) (dashed curves) and
exact FSR~(\ref{eq:FSR_exact}) (solid curves). Note that the latter
curves as well as the horizontal solid curves are identical to the solid
lines shown in Fig.~\ref{fig:1}.}
\label{fig:5}
\end{figure*}

The simplest way to improve the B2 approximation is to apply the
Pad\'{e} approximant to the second-order Born series in
Eq.~(\ref{eq:FSR_B2_c}). Applying the Pad\'{e} approximant of order
$[1/1]$ to this series one finds, instead of Eq.~(\ref{eq:alpha_B2}),
\begin{equation}
\alpha =
\alpha_{\mathrm{RPA}} \, \gamma^{1/2}
\left[ 1 - \frac{\pi}{2\gamma}\,Z_{1}\chi^{2}\,g(u) \right]^{-1/2}.
\label{eq:pade1}
\end{equation}
It is now important to trace the basic features of Eq.~(\ref{eq:pade1})
compared to the standard B2 approximation given by
Eq.~(\ref{eq:alpha_B2}). For a positive ion ($Z_{1}>0$) and at
$\chi^{2}\gg 1$ assuming that $\alpha/\alpha_{\mathrm{RPA}}$ increases
with $\chi$ (see Figs.~\ref{fig:1}--\ref{fig:4}) we get
\begin{equation}
\frac{\alpha}{\alpha_{\mathrm{RPA}}} \simeq
\frac{\pi C_{0}}{4\gamma} \, Z_{1}\chi
+ \sqrt{\gamma+\left(\frac{\pi C_{0}}{4\gamma}\,Z_{1}\chi\right)^{2}} \simeq
\frac{\pi C_{0}}{2\gamma} \, Z_{1}\chi.
\label{eq:pade2}
\end{equation}
In the case of a negative ion ($Z_{1}<0$) it is expected that the ratio
$\alpha/\alpha_{\mathrm{RPA}}$ decreases with $\chi$ (see
Figs.~\ref{fig:1}--\ref{fig:4}). Therefore, the solution of
Eq.~(\ref{eq:pade1}) must behave as $\alpha/\alpha_{\mathrm{RPA}}\simeq
A(Z_{1})/\chi$ at $\chi^{2}\gg 1$, where $A$ is independent of $\chi$
but depends on $Z_{1}$. The constant $A$ is then extracted from the
transcendental equation $A^{2}g(1/A)=\xi\equiv 2\gamma^{2}/\pi\vert
Z_{1}\vert$. In general the quantity $A$ behaves as $A\sim\xi^{\beta}$,
where the numerical constant $\beta$ varies between $1/3 \leqslant \beta
\leqslant 1$ with increasing $\xi$. The asymptotic solutions of
Eq.~(\ref{eq:pade1}) should be compared with
Eqs.~(\ref{eq:alpha_B2_asymptotic1}) and
(\ref{eq:alpha_B2_asymptotic2}). It is seen that the Pad\'{e}
approximant to the Born series in Eq.~(\ref{eq:FSR_B2_c}) at
$\chi^{2}\gg 1$ increases systematically the screening parameter both
for positive and negative ions. Consequently, as discussed above an
improvement of the B2 approximation is expected for any negative ion as
well as for a positive ion with lowest charge state $Z_{1}=+1$. For
$Z_{1}=+2$ and $+3$ it is expected that the Pad\'{e} approximant to the
Born series makes the agreement with the exact treatment even worse. As
an example we demonstrate these features in Fig.~\ref{fig:5}, where the
numerical solutions of the approximate Eq.~(\ref{eq:pade1}) are compared
with the exact values for the Yukawa potential. Comparing this figure
with Fig.~\ref{fig:1} one concludes that the Pad\'{e} approximant
essentially improves the agreement between the B2 approximation and the
exact results for any negative ion and in the whole interval of
$r_{\mathrm{s}}$. Such an improvement is also clearly visible for
$Z_{1}=+1$, while in the case of $Z_{1}= +2$, $+3$ the perturbative
approach strongly deviates from the self-consistent treatment based on
the FSR.

\end{document}